\documentstyle[aps,multicol]{revtex}

\newcommand{\al}{\alpha}
\newcommand{\ep}{\epsilon}
\newcommand{\th}{\theta}
\newcommand{\et}{\eta}
\newcommand{\rs}{{\cal R}}
\newcommand{\ls}{{\cal L}}

\input{psfig}

\draft
\begin{document}

\title{Scenarios of domain pattern formation in a reaction-diffusion
  system} \author{C. B. Muratov} \address{Department of Physics, Boston
  University, Boston, Massachusetts 02215} \author{V. V. Osipov}
\address{Department of Theoretical Physics, Russian Science Center\\ 
  ``Orion''\\ 2/46 Plekhanov St., Moscow 111123, Russia} \date{\today}
\maketitle
\begin{abstract}
  We performed an extensive numerical study of a two-dimensional
  reaction-diffusion system of the activator-inhibitor type in which
  domain patterns can form. We showed that both multidomain and
  labyrinthine patterns may form spontaneously as a result of Turing
  instability. In the stable homogeneous system with the fast inhibitor
  one can excite both localized and extended patterns by applying a
  localized stimulus.  Depending on the parameters and the excitation
  level of the system stripes, spots, wriggled stripes, or labyrinthine
  patterns form. The labyrinthine patterns may be both connected and
  disconnected. In the the stable homogeneous system with the slow
  inhibitor one can excite self-replicating spots, breathing patterns,
  autowaves and turbulence.  The parameter regions in which different
  types of patterns are realized are explained on the basis of the
  asymptotic theory of instabilities for patterns with sharp interfaces
  developed by us in Phys. Rev. E. {\bf 53}, 3101 (1996).  The dynamics
  of the patterns observed in our simulations is very similar to that of
  the patterns forming in the ferrocyanide-iodate-sulfite reaction.
\end{abstract}


\begin{multicols}{2}

\section{introduction}

Pattern formation and self-organization are among the most fascinating
phenomena in modern science that are observed in physical, chemical,
and biological systems of very different nature (for the books and
recent reviews on this subject see
\cite{nicolis,field,vasiliev,murray,mikhailov,lengyel93,%
cross93,kapral,ko:irreversible,ko:ufn90,ko:ufn89,ko:book,%
niedernostheide}, where a lot of references to the original works can
also be found).  As a rule, self-organization is associated with Turing
instability of the homogeneous state in the nonequilibrium systems and
spontaneous formation of patterns (dissipative structures) in them as
the excitation level of the system (or some other parameter) is varied
\cite{nicolis,field,vasiliev,murray,mikhailov,lengyel93,%
cross93,kapral,ko:ufn90,ko:book}.  At the same time, when the
homogeneous state of such a system is stable, by applying a
sufficiently strong perturbation one can excite static, pulsating, and
traveling patterns including solitary patterns --- autosolitons (AS)
\cite{ko:irreversible,ko:ufn89,ko:book,niedernostheide,bode95}.

Consider, for example, chemical patterns forming in the
ferrocyanide-iodate-sulfite (FIS) reaction in a gel reactor
\cite{lee:sci93,lee:ln94,lee:pre95}.  In a typical experimental setting
a pattern may form spontaneously as a result of the instability of the
homogeneous state or can be excited by a short localized external
perturbation of the system. The patterns forming in both these cases do
not have any qualitative differences.  They are essentially the domains
of high and low concentrations of certain substance separated by
relatively sharp walls. The pattern may have sophisticated geometry and
in general may show very complicated spatio-temporal behavior. The
properties of the patterns do not significantly depend on whether the
system is monostable or bistable.

It appears that domain patterns forming in very different systems may in
fact have many common features. This has recently been noticed in the
case of static domain patterns forming both in equilibrium and
nonequilibrium systems, where a certain set of domain shapes, such as
spots, stripes, multidomain and labyrinthine patterns, and the
transitions between them has been observed \cite{seul95}. The same
conclusion can be extended to the dynamic patterns in nonequilibrium
systems. Indeed, traveling, pulsating, self-replicating, and
stochastically oscillating patterns are observed in the systems as
diverse as autocatalytic reactions
\cite{kapral,lee:sci93,lee:ln94,lee:pre95}, semiconductor and gas plasma
\cite{ko:ufn90,ko:ufn89,ko:book,niedernostheide,bode95,ammelt93}, or
premixed flames \cite{gorman94:order,gorman94:hopping,gorman94:chaos}.
All this suggests that there exists a universality class of the
nonequilibrium systems in which pattern formation and self-organization
scenarios are essentially the same.

Another important question raised by the experiments is to identify the
totality of possible types of patterns and their behaviors in the
systems under consideration, and to understand the requirements the
system should meet in order to be able to produce one type of pattern or
the other.

In the present paper we will study a model which is a typical
representative of the pattern-forming systems whose phenomenology was
discussed above. We will show that the type of patterns that form in
this model is determined mainly by the relationship between the
characteristic length and time scales of the systems and the way the
system is excited. We will also show that by changing only these length
and time scales and choosing an appropriate form for the external
stimulus one can make the system form practically all kinds of patterns,
both static and dynamic, that are observed in the experiments.

Our paper is organized as follows: in Sec. II we discuss the physical
mechanisms of pattern formation phenomena in reaction-diffusion systems
of the activator-inhibitor type, using a combustion model as an example,
and introduce a simple model which we study numerically; in Sec. III we
present the results of a systematic numerical study of the
reaction-diffusion model and give qualitative explanations to the
effects seen; in Sec IV we use the general asymptotic theory of
instabilities developed by us in Ref.  \cite{mo1:pre96} to identify the
parameter regions in which one or the other type of patterns is observed
and compare these regions with the results of the simulations, we also
give more substantial quantitative explanations for the pattern
behaviors that are observed basing on the interfacial dynamics approach;
and finally, in Sec. V we compare our results with various experiments
and draw conclusions.

\section{the model}

The model which describes the phenomenology of pattern formation in
many nonequilibrium systems is a pair of reaction-diffusion systems of
the activator-inhibitor type
\begin{equation} \label{1}
  \tau_\th {\partial \th \over \partial t} = l^2 \Delta \th - q(\th,
  \et, A),
\end{equation} 
\begin{equation} \label{2} 
  \tau_\et {\partial \et \over \partial t} = L^2 \Delta \et - Q(\th,
  \et, A),
\end{equation} 
where $\th$ is the activator, $\et$ is the inhibitor, $l$ and $L$ are
the characteristic length scales, and $\tau_\th$ and $\tau_\et$ are
the characteristic time scales of the activator and the inhibitor,
respectively, $q$ and $Q$ are certain non-linear functions, and $A$ is
the bifurcation parameter. Equations (\ref{1}) and (\ref{2}) have been
extensively used to study pattern formation in various nonequilibrium
systems. In particular, they describe electron-hole and gas plasma,
various semiconductor, superconductor, and gas-discharge structures,
systems with uniformly generated combustion material
\cite{ko:ufn90,ko:ufn89,ko:book,niedernostheide,bode95}; chemical
reactions with autocatalysis and cross-catalysis
\cite{field,lengyel93,kapral}; models of morphogenesis and population
dynamics in biology \cite{murray}.  Some systems with phase
transitions, such as diblock copolymer blends and ferroelectric
semiconductors, are also described by equations which can be reduced
to Eqs.  (\ref{1}) and (\ref{2}) \cite{ohta86,ohta90,mamin94}.

Pattern formation in the systems under consideration is associated with
a positive feedback of the activator $\th$ which results in
``self-production of the activator substance'', this process of
self-production is controlled by the inhibitor $\et$ that suppresses the
growth of the activator. It is these two competing processes that give
rise to different kinds of patterns in these systems.

The meaning of the variables $\th$ and $\et$ can be most easily
understood for the system with uniformly generated combustion material
\cite{ko:ufn89,ko:book}. Consider combustion process in the flow reactor
consisting of a chamber placed between the two porous slabs which are
being cooled. The mixture of fuel and oxidizer is pumped through the
narrow reactor region between the slabs where it is ignited.
Phenomenologically, this system may be described by the equations for
the mass diffusion and the heat conductance which include the reaction
terms, averaged over the thickness of the reactor region. If $n$ is the
concentration of the fuel and $T$ is the temperature of the mixture,
these equations have the form:
\begin{equation} \label{burn:i} 
  {\partial n \over \partial t} = D \Delta n + G(n) - R(n, T),
\end{equation}
\begin{equation} \label{burn:a} 
  c \rho {\partial T \over \partial t} = \kappa \Delta T + E R(n, T) -
  W(T),
\end{equation}
where $G = (n_0 - n)/ \tau_n$ is the rate of the fuel supply, $R(n, T) =
\alpha n \exp(-E_a / T)$ is the reaction rate, $E_a$ is the activation
energy, $E$ is the reaction heat and $\alpha$ is a coefficient; $W(T) =
c \rho (T - T_0)/ \tau_T$ is the heat removed from the mixture, $\rho$
is the density, $c$ is the heat capacity of the mixture, $T_0$ is the
temperature of the environment; $D$ and $\kappa$ are the diffusivity and
the thermal conductivity, respectively, $\tau_n$ and $\tau_T$ are time
constants, and $\Delta$ is the two-dimensional Laplacian. In this system
the activator is the temperature $T$, and the inhibitor is the
concentration $n$. Suppose that the system is initially in the
low-temperature state. Now, if a localized region of the mixture is
heated up, the rate of the reaction in that region will rapidly
increase, thus producing more heat and igniting the neighboring areas.
However, this process cannot go forever, since as the reaction proceeds,
the fuel is being used up, which in turn decreases the rate of the
reaction. Thus, a positive feedback is realized with respect to the
temperature and the negative feedback with respect to the concentration.
If we introduce the variables $\th = T / E_a$, $\et = n/n_0$, $l =
\sqrt{\kappa \tau_T / c \rho}$, $L = \sqrt{D \tau_n}$, and $A = \al n_0
E \tau_T / c \rho E_a$, we will arrive at Eqs. (\ref{1}) and (\ref{2}).

Pattern formation is most pronounced in chemical and biological systems
\cite{nicolis,field,vasiliev,murray,mikhailov,lengyel93,kapral,%
  lee:sci93,lee:ln94,lee:pre95} where the processes of self-production
of matter are responsible for it. However, in the real situation such
systems are extremely complicated. Nevertheless, it is often possible to
reduce the description of these systems to a pair of reaction-diffusion
equations of the activator-inhibitor type
\cite{nicolis,field,vasiliev,murray,mikhailov,lengyel93,kapral}.  A
particularly simple model of this kind is the ``cubic'' model, which is
described by Eqs. (\ref{1}) and (\ref{2}) with
\begin{equation} \label{q}
  q = \th^3 - \th - \et,
\end{equation} 
\begin{equation} \label{Q} 
  Q = \th + \et - A.
\end{equation}
This model will be studied numerically and analytically in the
subsequent sections.

Kerner and Osipov showed that the properties of the patterns and pattern
formation scenarios in the systems described by Eqs. (1) and (2) are
chiefly determined by the parameters $\ep = l / L$ and $\al = \tau_\th /
\tau_\et$, and the shape of the nullcline of Eq. (1) for the activator
\cite{ko:ufn90,ko:ufn89,ko:book}. In many cases, including the cubic
model and the model described by Eqs. (\ref{burn:i}) and (\ref{burn:a}),
this nullcline is N-shaped (see Fig.~\ref{null}).  In these systems
static, pulsating, and traveling patterns may form at different values
of the parameters $\ep$ and $\al$.

When $\al \ll 1$ and $\ep \gtrsim 1$, that is when the inhibitor is
slower and shorter-ranged than the activator, only traveling patterns
may exist. In the limit $\ep \rightarrow \infty$ (or, more precisely,
for $L = 0$) the properties of such traveling patterns were studied in
detail in both one-dimensional and higher-dimensional cases (see, for
example, \cite{field,vasiliev,mikhailov,kapral} and references
therein). In the other limiting case $\ep \ll 1$ and $\al \gtrsim 1$,
that is when the inhibitor is long-range and fast compared to the
activator, only static patterns may form
\cite{ko:ufn90,ko:ufn89,ko:book}.  These patterns are essentially the
domains of high and low activator values separated by the narrow walls
(interfaces) whose width is of order $l$.  Traveling, static, and
pulsating domain patterns may form when both $\ep \ll 1$ and $\al \ll
1$.

Let us elucidate the physics of the formation of static, traveling, and
pulsating patterns, including the simplest patterns --- AS (Fig.
\ref{as}), using the combustion system described by Eqs. (\ref{burn:i})
and (\ref{burn:a}).  First of all, it is clear that if both the
characteristic time and length scales of the variation of the
concentration $n$ are much smaller than the characteristic scales of the
variation of the temperature $T$, no patterns will form. Indeed, if a
localized region of the system is ignited, all the fuel will burn down
very fast in that region and the flame will not be able to propagate, so
after a short time after removing the heat source the flame will
extinguish. Different situation is realized if the diffusivity of the
fuel is much smaller than the thermal diffusivity and the characteristic
time scale of the temperature relaxation $\tau_T$ due to the heat
exchange between the mixture and the porous slabs is much longer than
the characteristic time scale $\tau_n$ of the concentration variation
determined by the rate of the fuel supply. Then the conditions $\al \ll
1$ and $\ep \gtrsim 1$ may be satisfied and the patterns in the form of
traveling flames may be excited in the system. The existence of
traveling patterns is due to the fact that because of high heat
conductance the region of size of order $l = \sqrt{\kappa \tau_T / c
  \rho}$ around the flame is heated up and ignited. As a result, the
released heat ignites the neighborhood and so on. This leads to the
formation of a flame front moving with the speed $v \simeq l / \tau_T$
[Fig. \ref{as}(a)].  The fuel after the front burns down, so the front
is followed by the back separated from the front by the distance of
order $v \tau_n \gg l$. Because of the external supply the fuel
replenishes at the distances of order $v \tau_n$ away from the back.

If the diffusivity of the fuel is much larger than the thermal
diffusivity of the mixture, the condition $\ep \ll 1$ and $\al \gtrsim
1$ may be satisfied. Then a traveling flame front will stop since high
diffusion of the fuel will result in the decrease of the fuel
concentration ahead of the front and lead to the formation of the static
pattern in the form of a flame cell [Fig. \ref{as}(b)]. The existence of
such static pattern is due to the fact that the flame is maintained by
the diffusive influx of the fuel from the neighborhood, where it is
constantly supplied through the porous slabs. It is clear that when both
$\ep \ll 1$ and $\al \ll 1$, pulsating flames, or more complex dynamic
patterns, may form in the system [Fig. \ref{as}(c)]. The cell may first
expand, but as it cools down and the fuel is used up in it, the front
may stop and start traveling back \cite{ko:ufn89,ko:book}.

From the physical considerations above follows that in order for
patterns to be able to form, either $\ep$ or $\al$ has to be small.
Kerner and Osipov suggested the classification scheme for the
reaction-diffusion systems of the activator-inhibitor type based on
the relationship between the values of $\ep$ and $\al$
\cite{ko:irreversible,ko:ufn90,ko:ufn89,ko:book}.  According to this
scheme, a system with N-shaped nullcline of the equation for the
activator is called $\Omega$N system if $\al \ll 1$ and $\ep \gtrsim
1$; KN system, if $\ep \ll 1$ and $\al \gtrsim 1$; or K$\Omega$N
systems, if both $\al \ll 1$ and $\ep \ll 1$. Accordingly, only
traveling waves (autowaves) may form in $\Omega$N systems, only static
patterns in KN systems, and all kinds of patterns in K$\Omega$N
systems.

Clearly, the mechanisms of pattern formation in all reaction-diffusion
systems described by Eqs. (\ref{1}) and (\ref{2}) are essentially the
same as those discussed above. For example, similar processes lead to
the formation of static, traveling, and pulsating patterns in the form
of the regions of high temperature and low concentration of electrons in
the photo-generated electron-hole plasma heated in the process of Auger
recombination \cite{ko:ufn89,ko:book,ko:jetp85}. For this reason we may
use the simple ``cubic'' model described by Eqs. (\ref{1}) and (\ref{2})
with (\ref{q}) and (\ref{Q}) in all our numerical simulations. Depending
on $\al$ and $\ep$, this system will pertain to one of the three classes
mentioned above. In this paper we will concentrate on the case when the
system under consideration is either KN or K$\Omega$N system.

\section{Scenarios of pattern formation: results of the simulations}

If we choose $L$ and $\tau_\et$ as the units of length and time,
respectively, we will write the equations describing the cubic model in
the form
\begin{equation} \label{act}
  \al {\partial \th \over \partial t} = \ep^2 \Delta \th - \th^3 + \th +
  \et, \end{equation}
\begin{equation} \label{inh}
  {\partial \et \over \partial t} = \Delta \et - \th - \et + A.
\end{equation}

We performed numerical simulations of Eqs. (\ref{act}) and (\ref{inh})
in two dimensions in a wide range of the parameters $\ep$, $\al$, and
$A$. As we mentioned earlier, either $\ep$ or $\al$ has to be small in
order for patterns to be able to form. The case of $\Omega$N systems
was extensively studied by many authors
\cite{field,vasiliev,mikhailov,kapral} and will not be studied
here. We will concentrate on the case $\ep \ll 1$ and arbitrary $\al$.

The presence of two very different length scales caused by the smallness
of the parameter $\ep$ makes the numerical simulations of Eqs.
(\ref{act}) and (\ref{inh}) rather difficult. The simulations were
performed on the massively parallel supercomputer. An explicit second
order finite difference scheme was used to discretize the equations. In
order to accelerate the algorithm, different grid spacings were used for
$\th$ and $\et$. The boundary conditions were neutral or periodic. The
grid spacing was chosen in such a way that a typical front of a pattern
contained about 8 to 10 points. The decrease of the grid spacing by a
factor of two resulted in the difference of the distributions of $\th$
and $\et$ by a few percent. No noticeable effects on the dynamics were
observed.

Before discussing the results of the simulations, let us make a few
comments about the system of Eqs. (\ref{act}) and (\ref{inh}). First of
all, it is easy to see that these equations are invariant with respect
to the transformation
\begin{equation} \label{trans}
  \th \rightarrow -\th, ~~~\et \rightarrow -\et, ~~~A \rightarrow -A,
\end{equation}
so one only needs to study the parameter region where $A < 0$. The
system under consideration is monostable for all values of $A$. The
homogeneous state $\th = \th_h$ and $\et = \et_h$ with
\begin{equation} \label{hom}
  \th_h = - | A |^{1/3}, ~~~\et_h = |A|^{1/3} ( 1 - |A|^{2/3} )
\end{equation}
is stable for $A < A_0 = -1/3\sqrt{3} \simeq -0.19$, where $A_0$ is
the value of $A$ at which $\th_h = \th_0$ and $\et_h = \et_0$, the
point on the nullcline of Eq. (\ref{act}) at which $q'_\th = 0$ (see
Fig.  \ref{null}). It is easy to see
\cite{ko:irreversible,ko:ufn90,ko:ufn89,ko:book} that for $\al < 1$
the homogeneous state becomes unstable with respect to the uniform
oscillations (Hopf bifurcation) with the frequency
\begin{equation} \label{Aom}
  \omega_0 = \left( {1 - \al \over \al} \right)^{1/2}~~{\rm at }~~ A >
  A_\omega = - \left( { 1 - \al \over 3 } \right)^{3/2},
\end{equation}
whereas for $\ep < 1$ it destabilizes with respect to the fluctuations
with the wave vector $k = k_0$ and zero frequency (Turing bifurcation)
\begin{equation} \label{Ac}
  k_0 = \left( { 1 - \ep \over \ep } \right)^{1/2} ~~{\rm at}~~ A > A_c
  = - \left( { 1 - \ep \over \sqrt{3} } \right)^3.
\end{equation}
Notice that for $\ep \rightarrow 0$ or $\al \rightarrow 0$ we have $A_c
\rightarrow A_0$ or $A_\omega \rightarrow A_0$, respectively. Also
notice that the homogeneous state is always stable when $\ep > 1$ and
$\al > 1$. It is the considerations of stability of the homogeneous
state that actually lead Kerner and Osipov to divide the system
described by Eqs. (\ref{1}) and (\ref{2}) into $\Omega$N, KN, and
K$\Omega$N systems \cite{ko:ufn90,ko:ufn89,ko:book}. Similar
classification of pattern-forming systems of different nature, including
the hydrodynamic systems, was proposed later by Cross and Hohenberg
\cite{cross93}.

Most of our simulations were performed at $\ep = 0.05$, what can be
considered reasonably small. In the first simulations we studied the
Turing instability. The boundary conditions in these simulations were
periodic. The initial condition was taken in the form of the
homogeneous state plus small random noise. The system then evolved for
a considerably long time at $A = -0.1$ and $\al = 0.5$ when the
homogeneous state is unstable with respect to Turing instability but
stable with respect to the oscillatory instability. The process of
formation of static Turing pattern is shown in Fig. \ref{turing}. In
the early stage ($t = 6$ and $t = 9$ in Fig. \ref{turing}) the system
nucleates some random distribution of the activator and the inhibitor.
At the intermediate stage ($t = 16$) the pattern transforms into a
number of domains with sharp walls, at this point the domains may have
irregular shapes and domain fusion frequently occurs; at the late
stage ($t = 51$ and $t = 210$) the domains rearrange so that their
shape becomes more regular, and a lot of smaller domains die through
overcrowding. One can see that in the end the static Turing
(multidomain) pattern consists of many disconnected domains with sharp
walls. Most of them look like slightly distorted circular domains of
different sizes, although some are more stripe-like. The pattern is
metastable, upon longer runs a small portion of domains may
occasionally disappear or change their geometry, but its overall
appearance remains the same. The interaction between the neighboring
domains appears to be repulsive, so one should expect an ideal
hexagonal pattern of circular domains of certain radius to be the most
stable one. This pattern reminds of the ordered cellular flame pattern
observed in the combustion experiments \cite{gorman94:order} and
ordered pattern of current filaments in the gas-discharge experiments
\cite{ammelt93}.

In the next simulation the initial condition is taken the same, but
the parameters now are $A = -0.1$ and $\al = 0.05$, so that the
homogeneous state is unstable both with respect to the Turing
instability and the oscillatory instability. The evolution of the
system toward the static labyrinthine pattern in this case is shown in
Fig. \ref{unstab}. The early stage of the formation of the pattern ($t
= 0.6$ and $t = 0.9$) is the same as in the previous case, but once
the domains form they start to oscillate (breathe), so the high
activator value domains invade almost all the space in the time
interval from $t = 1.2$ to $t = 2.1$, but then recede, and the process
continues. For this value of $\al$ the pattern finally stabilizes, and
eventually a static labyrinthine pattern forms. Notice that both the
multidomain pattern which formed at $t = 955$ in Fig.  \ref{turing}
and the labyrinthine pattern which formed at $t = 9.2$ in
Fig. \ref{unstab} are perfectly good Turing patterns formed as a
result of Turing instability, so there is no qualitative difference
between the two. The form of the Turing pattern in any particular
situation must therefore depend on the history of getting the
homogeneous system into the unstable state. The form of the Turing
pattern may also strongly depend on the small local inhomogeneities
always present in real systems
\cite{ko:ufn90,ko:ufn89,ko:book}. Notice that both the multidomain and
labyrinthine Turing patterns were observed in chemical experiments
\cite{lengyel93,kapral,lee:pre95}.

Our simulations show that for sufficiently small $\al$ the uniform
self-oscillations of the homogeneous state will always set in upon the
destabilization of the homogeneous state even if the system is
unstable with respect to the Turing instability. Specifically, for
$\ep = 0.05$ this will happen, if $\al < 0.02$. For larger values of
$\al$ the static Turing pattern will always form, although the
oscillations of the pattern may last a long time.

In the next series of the simulations we investigate the formation of
patterns when the homogeneous state of the system is stable. In all
these simulations the boundary conditions were neutral.

It is well known that when $\al \ll 1$ and $\ep \gtrsim 1$ ($\Omega$N
systems) Eqs.  (\ref{1}) and (\ref{2}) admit solutions only in the
form of traveling waves (autowaves)
\cite{field,vasiliev,mikhailov,kapral,ko:ufn90,ko:book,%
ortoleva75,casten75}, and when $\ep \ll 1$ and $\al \gtrsim 1$ (KN
systems) they admit only solutions in the form of static patterns, the
simplest of which are AS in the form of solitary spots and stripes of
high or low values of the activator surrounded by the ``sea'' of low
or high values of the activator, respectively (``hot'' and ``cold''
AS) \cite{ko:ufn89,ko:book}. Notice that because of the monostability
of the considered system the radius of a spot or the width of a stripe
cannot be greater than certain value of order one
\cite{ko:ufn89,ko:book}. Also note that because of the symmetry given
by Eq. (\ref{trans}), in the system under consideration we only need
to consider the behavior of hot patterns.

When the homogeneous state of the systems is stable, the patterns may
be excited by means of sufficiently strong external stimulus
\cite{ko:ufn89,ko:book}. According to the general qualitative theory,
the excitation level $A$ of the system must be greater than certain
threshold value $A_b$ in order for AS to be able to form
\cite{ko:ufn90,ko:ufn89,ko:book}. It is possible to show that in the
limit $\ep \rightarrow 0$ the value of $A_b = -1$ in the considered
system.

First we will consider KN systems. Our numerical simulations show that
when $A$ is close to $A_b$, the initial condition in the form of the
homogeneous state plus a hot spot of size of order several $\ep$ evolves
into an AS in the form of a localized static radially-symmetric spot.
For $\ep = 0.05$ this happens if $A_b^{(2)} < A < A_{c2}^{(2)}$, where
$A_b^{(2)} \simeq -0.72$ and $A_{c2}^{(2)} \simeq -0.55$. If the initial
condition is taken in the form of the homogeneous state plus a hot
stripe several $\ep$ wide, it will evolve into a static stripe if
$A_b^{(1)} < A < A_{c2}^{(1)}$, where $A_b^{(1)} \simeq -0.74$ and
$A_{c2}^{(1)} \simeq -0.55$ (the value of $A_{c2}^{(1)}$ obtained from
the simulations is rather crude since the destabilization of the stripe
may be incredibly slow).

If the value of $A$ is increased from $A_b^{(2)}$ to $A_{c2}^{(2)}$, the
radius of the spot will grow. However, at certain radius corresponding
to the value of $A = A_{c2}^{(2)}$ the spot becomes unstable with
respect to the radially-nonsymmetric distortions of its walls
\cite{ko:ufn89,ko:book,mo1:pre96,ohta89}. Qualitatively, this means that
a ``burning spot'' tends to ignite the neighboring regions, but since
its radius is bounded from above, the only way it can grow in size is by
elongation. Precisely this phenomenon was observed in the simulations.
For $A < A_{c3}^{(2)} \simeq -0.46$ the spot elongates and transforms
into a stripe. If $A > A_{c3}^{(2)}$, the tips of the growing stripe
become further unstable, leading to the tip splitting and the formation
of a labyrinthine pattern. This effect is seen in the simulation of Fig.
\ref{nosplit}, where $A = -0.4$, which is not much greater than
$A_{c3}^{(2)}$. There an almost radially-symmetric spot at $t = 0$
elongated and transformed into a dumbbell at $t = 63$ and then further
destabilized into a more complex shape at $t = 98$. The process of tip
splitting resulted in a more and more complicated pattern at $t = 162$
and $t = 292$, until the resulting labyrinthine pattern reached the
system boundaries and stopped growing at $t = 985$.  Similar structures
were also observed in the simulations in Refs.
\cite{hagberg:prl94,goldstein96} and in chemical reactions
\cite{lengyel93,kapral,lee:pre95}. The labyrinthine pattern that
formed in the end is a collection of stripe-like domains which are all
connected. Notice that in this simulation $\al = 0.2$. However, the
smallness of the value of $\al$ does not affect the simulation results
as long as $\al \gg \ep$. On the other hand, this choice of $\al$
significantly accelerates the simulations.

In the next simulation we took $A = -0.25$, which is further away from
$A_{c2}^{(2)}$ and closer to $A_c$. The initial conditions were taken
in the form of the homogeneous state plus a piece of a curved
stripe. One can see (Fig. \ref{corr}) that initially ($t = 22$) the
tips of the pattern grow faster and start splitting ($t = 87$). One
can also see that the stripe itself start to wriggle, and fingers
spring out of the regions with the highest curvature ($t =
118$). However, in contrast to the previous case, some of the portions
of the growing labyrinthine pattern detach themselves from the body of
the pattern. As the time passes, more and more portions become
detached. In the end the labyrinthine pattern that fills the whole
system at $t = 322$ consists of 5 disconnected pieces. Notice the
great similarity between this pattern and the labyrinthine pattern
formed as a result of Turing instability in Fig. \ref{unstab}. Also
notice that the disconnected pattern is more likely to form even for
$A$ not very different from $A_{c2}^{(2)}$, if $\al$ is smaller. This
is because at small $\al$ the dynamics of the pattern becomes
oscillatory and the labyrinthine pattern may form as a result of
self-replication of spots, which will be discussed below.

The same picture can be observed, if the initial conditions are taken
in the form of a twisted stripe running across the system, if $A$ is
big enough. In general, for $A$ considerably greater than $A_b$ any
localized initial condition will lead to the formation of the
disconnected labyrinthine pattern. However, if these boundary
conditions are used with the values of $A$ closer to $A_b$, a wriggled
stripe pattern will form in the system. This process is related with
the fact that a stripe may become unstable with respect to wriggling
of the stripe as a whole while being stable with respect to fingering
when $A_{c2}^{(1)} < A < A_{c1}^{(1)}$
\cite{ko:ufn89,ko:book,mo1:pre96,ohta89}. For $\ep = 0.05$ the value
of $A_{c1}^{(1)}$ obtained from the simulations is $A_{c1}^{(1)}
\simeq -0.32$. The evolution of the stripe at $A = -0.45$ is shown in
Fig.  \ref{wrig}. One can see that the stripe gets more and more
wriggled without fingering for a long time. Only when the curvature of
some portion of the stripe becomes sufficiently high, a finger springs
out ($t = 1418$). Notice that fingering also occurs at the points
where the stripe is attached to the boundary. At these points the
curvature of the stripe is high as well.

Up to now we considered the pattern formation in the stable KN system in
which the inhibitor is fast. According to our simulations, the time
scale of the inhibitor variation does not affect all the results above
when $\al \ll 1$ but $\al \gg \ep$. In K$\Omega$N systems, in which the
inhibitor is slow enough, pattern formation scenarios will be
qualitatively different.  Figure \ref{split} shows the evolution of the
system at $A = -0.4$, but $\al = 0.015$ with the initial condition in
the form of an almost radially-symmetric spot. In contrast to the
simulation of Fig. \ref{nosplit} which was performed for the same value
of $A$ and with the same initial conditions, instead of transforming
into a dumbbell the spot splits into two in the course of its evolution
($t = 3.3$). The spots that form split in turn into four ($t = 5.2$).
This process of self-replication of spots continues until the whole
system is filled with the multidomain pattern (not shown in the figure),
which may stabilize or transform into a synchronously pulsating
(breathing) pattern. Notice that self-replication of spots was observed
in the chemical experiments \cite{lee:ln94,lee:pre95} and in the
simulations \cite{hagberg:prl94}. A similar phenomenon also seems to
occur in the chaotic cellular flames \cite{gorman94:chaos}.

Figure \ref{osc} shows the value of $\et$ in the center of the system in
which the synchronously pulsating multidomain pattern formed as a result
of spot self-replication. One can see that the multidomain pattern forms
at relatively short times ($t \lesssim 5$ for the system $10 \times
10$), and after that the oscillation of the pattern as a whole starts.
Some restructuring of the pattern occurs later on, what results in the
changes of the oscillations amplitude and the pattern's geometry. At
times $t \gtrsim 120$ the pattern's oscillations had synchronized and no
changes in the oscillations amplitude nor in the pattern's geometry were
observed in the longer runs.

The mechanism of self-replication can be seen from Fig. \ref{sd}, where
a single self-replication event is shown in detail. One can see that
self-replication is determined by the two processes: radially-symmetric
pulsations of the spot's radius and aperiodic growth of the
non-symmetric distortion. At the beginning the spot expands as a whole
($t = 0.6$), but at the same time the non-symmetric distortion builds up
($t = 1.4$). Then the spot starts to shrink in the course of the
radially-symmetric pulsations, so the connection between its right and
left portions gets torn at $t = 2.1$. At $t = 2.6$ there are two spots
looking just like the one at $t = 0$ in the system. Notice that for
smaller values of $A$ a single self-replication act may take more than
one pulsation period.

When the value of $\al$ is smaller, the process of self-replication of
domains may become stochastic, producing a kind of turbulence (Fig.
\ref{turb}). In the simulation of Fig. \ref{turb} ($\al = 0.01, A =
-0.4$) the initial condition in the form of a small domain initially
grows in size, but at $t = 1.2$ local breakdown occurs in its center, so
that the domain transforms into an annulus. The annulus then splits in
turn into several smaller domains ($t = 5.3$) which engage into
incessant stochastic motion. Each domain is self-replicating, but some
of the domains formed as a result of this process die as a result of the
collisions with the other domains, what causes the stochastization.
Another source of stochastization is the local breakdown which occurs,
if the domain size becomes too big. The number of domains in the system
changes randomly with time. Each domain is also moving as a whole. The
interaction between different domains (and the boundaries) is repulsive,
so the domain fusion is typically avoided. The turbulent pattern is
persistent and does not synchronize even after long times (Fig.
\ref{chaos}). It is observed only at sufficiently large values of $A$.
When $A$ is relatively small, the turbulent pattern usually collapses
into the homogeneous state after relatively short times. This kind of
turbulence was observed in the chemical \cite{lee:ln94,lee:pre95} and
combustion \cite{gorman94:chaos} experiments. Notice that the turbulence
that is observed in our simulations is different from the spiral
turbulence observed by Hagberg and Meron \cite{hagberg:prl94}. In our
simulations we never saw the nucleation of the spiral vortex pairs.

When the value of $\al$ is even smaller, a localized initial
perturbation transforms into an autowave. In the simulation of Fig.
\ref{waves}(a), in which $\al = 0.007, A = -0.3$, the domain expands,
and at $t = 1.1$ it transforms into an annulus which now remains
stable and continues to expand. This results in an autowave passing
through the system and annihilating when it reaches the boundaries. In
this case there is no repulsion between the autowaves. At these and
smaller $\al$ only autowaves form in the system, regardless of the
value of $A$. If a random initial condition is taken, spiral
turbulence typical of the excitable autowave media ($\Omega$N systems)
will form at $\al \ll \ep$ [Fig. \ref{waves}(b)]. Here the turbulent
pattern consists of a random arrangement of spiral vortices whose
positions are fixed in space. The spiral waves always annihilate upon
collision in this case. This is a well-known phenomenon in the
$\Omega$N systems (excitable media), for which $L = 0$
\cite{vasiliev,mikhailov,kapral}.

Before concluding this Section, let us mention two other simulations. In
the first [Fig. \ref{waves}(c)], for which $\al = 0.02$ and $A = -0.3$,
a localized initial perturbation results in a few replication-like acts,
but after that the pattern stabilizes into a static disconnected
labyrinthine pattern. In the second the value of $\ep = 0.2$ was taken
to be not very small (the other parameters are $\al = 1.$, and $A =
-0.1$), so that the system is away from the asymptotic regime $\ep
\rightarrow 0$ [Fig. \ref{waves}(d)]. One can see from Fig.
\ref{waves}(d) that the localized initial perturbation transforms into a
disconnected labyrinthine pattern in this case as well, so,
qualitatively, the effects observed in this Section are realized when
the value of $\ep$ is not very small.

The patterns which were described above are the only kinds of patterns
that were observed in the system under consideration. No other types of
patterns were observed in the simulations, no matter what initial
conditions or the parameters $\al$, $\ep$, and $A$ were used (of course,
there are ``cold'' patterns, but in view of Eq. (\ref{trans}) they are
equivalent to the ``hot'' patterns studied above). Thus, these patterns
constitute the totality of the pattern types of the considered system.

\section{domains of existence of different types of
  patterns and scenarios of pattern formation}

In this Section we will analyze the pattern formation scenarios observed
in the previous Section, give quantitative explanation for the parameter
regions in which different patterns form, and explain the
transformations of one type of pattern to the other on the basis of the
general asymptotic theory of instabilities for patterns with sharp
interfaces developed by us in Ref. \cite{mo1:pre96}, and on the basis of
the interfacial dynamics approaches developed in Refs.
\cite{ohta89,goldstein96,m:pre96}.

Our simulations of the spontaneous formation of Turing patterns
confirm the conclusion of the general qualitative theory that at the
threshold of Turing instability large-amplitude patterns should form
abruptly in the system \cite{ko:ufn90,ko:book}. According to
Eq. (\ref{Ac}), for small $\ep$ Turing instability occurs at $A \simeq
A_0 = -1/3 \sqrt{3} \simeq -0.19$, with respect to the fluctuations
with the wave vector $k_c \simeq \ep^{-1/2}$. One can see from
Fig. \ref{turing} that at early stages ($t = 16$) there are many
domains of small size, so one could naturally assume that at early
stages the domain sizes are determined by the wavelength of the
critical fluctuations, which is of order $\ep^{1/2}$. However, as can
be seen from Fig. \ref{turing}, at late stages the average size of the
domains becomes greater and in the end all domains have roughly the
same size.  This is not surprising since the domains of small size are
unstable because of the effect of the activator repumping
\cite{ko:ufn90,ko:book}. The thing is that because of its long-range
character, it is difficult for the inhibitor to react on such
variations of the activator that lead to the expansion of some of the
domains and the simultaneous shrinkage of their neighbors. This can be
seen from the estimate of the terms in the dispersion relation for the
fluctuations around the Turing pattern. For simplicity let us consider
a hexagonal arrangement of circular domains of radius $\rs_s$ with the
period $\ls_p$. Then the fluctuation which leads to the activator
repumping has the wave vector $k = \pi / \ls_p \gg 1$ for $\ls_p \ll
1$.  The term in the dispersion relation that causes the instability
is $\lambda_0 \simeq - \ep^2 / \rs_s^2$
\cite{ko:ufn89,ko:book,mo1:pre96}, whereas the inhibitor reaction
term, which has the stabilizing effect, is of order $\ep \ls_p$
\cite{ko:ufn90,ko:book}, and the values of $\ls_p$ and $\rs_s$ are of
the same order. Therefore, when $\rs_s$ is smaller than $\rs_b \sim
\ep^{1/3}$, this fluctuation will grow and lead to the expansion of
every second domain and collapse of the rest, what will result in the
increase of the pattern's period and the radius of the domains. In
other words, the domains will grow by eating their neighbors until
their size and the distance between them becomes of order $\ep^{1/3}$.

On the other hand, the domain radius cannot be greater than $\rs_{c2}
\sim \ep^{1/3}$ since at greater radii the domain becomes unstable with
respect to the non-symmetric deformations and either splits or
elongates.  The important thing, however, is that both $\rs_b$ and
$\rs_{c2}$ are much greater than $2 \pi / k_c \sim \ep^{1/2}$ for small
$\ep$, so the process of formation of Turing pattern must always consist
of two stages: initial domain forming and ripenening.

Notice that in the presence of small localized inhomogeneities the
process of formation of Turing pattern may be qualitatively different
\cite{ko:ufn89,ko:book}. A small localized domain may nucleate at the
inhomogeneity, but then as a result of the transverse instability of
its walls, which occurs when the domain radius becomes of order
$\ep^{1/3}$ \cite{mo1:pre96}, it will transform into a disconnected
labyrinthine pattern, if $\ep$ is not very small, or start to split
and replicate itself until the system is filled with the domains of
size of order $\ep^{1/3}$, if $\ep$ is very small
\cite{m:pre96}. These effects will occur when $\al \gg \ep$.

According to Eqs. (\ref{Aom}) and (\ref{Ac}), Turing instability is
the first if $\al > 2 \ep$ for small $\ep$. However, as we see from
the simulations, even for smaller values of $\al$, that is, when the
homogeneous state of the system is unstable with respect to both
Turing and oscillatory instability, static Turing patterns may persist
up to smaller values of $\al$. For $\ep = 0.05$ this happens down to
$\al \simeq 0.02$.  For these values of $\al$ one can see the
competition between the Turing patterns and the uniform
self-oscillations. For $\al < 0.02$ the uniform self-oscillations win,
and Turing patterns do not form. For $\al > 0.02$ the situation is
reverse. We did not observe the coexistence of Turing patterns and
uniform self-oscillations. It is interesting to note, however, that
the well-formed Turing pattern may be stable for even smaller values
of $\al$. Also, if there are small local inhomogeneities in the system
with $\al \sim \ep$, the system will nucleate localized domains before
reaching the instability, as in the case of $\al \gg \ep$, but then
the domains will self-replicate and a multidomain pattern will form in
the system. If the value of $\al$ is smaller, the inhomogeneities will
cause nucleation of guiding centers \cite{ko:ufn89,ko:book}.

Let us now turn to the patterns that are excited in the system with the
stable homogeneous state. Our first observation is that, as was expected
\cite{ko:ufn89,ko:book} and in agreement with the statements of Fife
\cite{fife:jcp76}, any localized initial perturbation at first
relatively quickly transforms into a state in the form of the domain
with sharp walls, which is the closest to the initial perturbation in
shape, and then this domain starts to evolve considerably slower
according to the equations of the interface dynamics. The characteristic
time scale for the domain to form is that of the activator, that is
$\al$, and the characteristic time scale of the interface motion is $\al
/ \ep$ \cite{m:pre96}, so one can see that as long as $\ep \ll 1$ the
latter time is much longer than the former. In this sense one could
think that at first the initial perturbation evolves into a closest in
shape stationary state, which then grows into a more complicated stable
pattern as a result of the instability of that state. In this process
the early stages of the formation of a complex pattern is determined by
the type of the critical fluctuation with respect to which that
stationary state loses its stability. Therefore, it is important to know
the form of possible stationary states and when they become unstable.

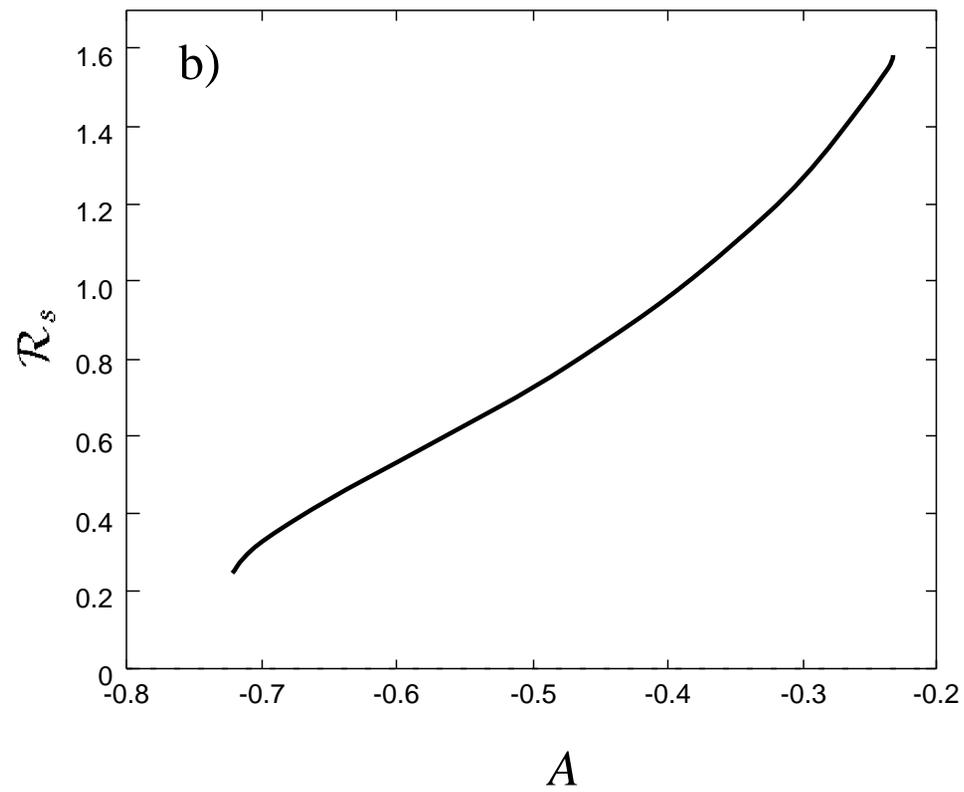
\begin{figure}
\caption{ \label{alar}
  The dependences of ${\cal L}_s$ on $A$ for the one-dimensional AS (a)
  and ${\cal R}_s$ on $A$ for the radially-symmetric AS (b) for $\ep =
  0.05$. Results of the numerical simulations of Eqs.
  (\protect\ref{act}) and (\protect\ref{inh}).}
\end{figure}

The simplest patterns in the considered system are static spots and
stripes \cite{ko:ufn89,ko:book,ohta89}.  They are indeed observed when
$A$ is sufficiently close to $A_b$, when a spot-like or stripe-like
initial perturbations are used, respectively. We performed numerical
simulations of the one-dimensional and radially-symmetric versions of
Eqs. (\ref{act}) and (\ref{inh}) and found the dependences of the
stripe's width $\ls_s$ and the spot's radius $\rs_s$ versus $A$ at $\ep
= 0.05$ (Fig. \ref{alar}). From these simulations one can see that the
solution in the form of a single static stripe exists at $A_{b}^{(1)} <
A < A_c$, where $A_b^{(1)} = -0.74$, whereas the solution in the form of
a single static spot exists when $A_b^{(2)} < A < A_d^{(2)}$, where
$A_b^{(2)} = -0.72$ and $A_d^{(2)} = -0.24 < A_c$. When $A_d^{(2)} < A <
A_c$, the local breakdown occurs in the spot's center, so the spot
transforms into an annulus. The thing is that for $\ep \ll 1$ the
distributions of the activator and the inhibitor outside the walls of
the spot are related via the equation of local coupling
\cite{ko:ufn90,ko:ufn89,ko:book,mo1:pre96}
\begin{equation}
  \label{lc}
  q(\th, \et) = 0.
\end{equation}
In other words, $\th$ and $\et$ lie on one of the stable branches of the
nullcline of Eq. (\ref{act}): $\th < \th_0$ in the cold region and $\th
> \th_0'$ in the hot region (Fig. \ref{null}). As the radius of the hot
spot grows the value of $\et$ in its center gets smaller, so at some
value of $A = A_d^{(2)} < A_c$ it reaches $\et_0'$, the point at which
the dependence $\th(\et)$ determined by Eq. (\ref{lc}) becomes singular,
so a sudden down-jump from one branch of the nullcline to the other
occurs, resulting in the formation of a new interface in the spot's
center and the transformation of the spot into an annulus.  Notice that
the process of local breakdown in the center of a spot and the formation
of an annulus in N systems was studied in detail in Refs.
\cite{ko:ufn89,ko:book}.  Also notice that the same mechanism is
responsible for the local breakdown in the center of a one-dimensional
stripe \cite{ko:ufn89,ko:book}. However, it does not occur in the
particular system we study.

In higher dimensions spots and stripes undergo instabilities leading to
the growth of certain deformations of their walls. Recently we developed
a general asymptotic theory of the instabilities of domain patterns in
arbitrary N systems \cite{mo1:pre96}. We have shown that the
instabilities are determined by the motion and the interaction of the
pattern's walls (interfaces). For sufficiently small $\ep$ one could use
the formulas obtained in Ref. \cite{mo1:pre96} and the dependences
$\ls_s(A)$ and $\rs_s(A)$ to determine the critical values of $A$ at
which one or another instability of the spots and stripes occur. The
parameters that enter those formulas for the considered system are
\begin{equation} \label{BZC} 
  B = 4, ~~~Z = \frac{2 \sqrt{2}}{3},~~~C = \frac{3}{2}.
\end{equation}
However, for $\ep = 0.05$ the agreement between the results of the
simulations and the predictions of Ref. \cite{mo1:pre96} is rather crude
(about 50\%). This is due to the fact that in the derivations of the
critical values of the domain sizes, which are typically of order
$\ep^{1/3}$, we used them as small parameters. However, because of the
slow $\ep$-dependence (1/3 power) this is not a very good assumption for
$\ep \simeq 0.05$.

Nevertheless, there is a way to calculate the critical values of $\ls_s$
and $\rs_s$ which agree with the results of the simulations with the
accuracy better than 5\%. To do this, we can use the dispersion
relations obtained in Ref. \cite{mo1:pre96} in the zeroth order of the
perturbation theory in the potential $V$, but not expanding in $\rs_s$,
or $\ls_s$, respectively, and keeping the value of $C$ evaluated at $A =
A_b$. The latter is because the critical fluctuations are localized in
the walls of the pattern and this is the way to take into account some
of the potential $V$. Having done this, for the stripes we have the
following dispersion relation
\begin{eqnarray} \label{d1}
  i \al \omega && + \ep^2 k^2 + \lambda_0 = \nonumber \\ && - {\ep B
    Z^{-1} \over 2\sqrt{C + k^2 + i \omega}} \left(1 \pm \exp (- \ls_s
  \sqrt{C + k^2 + i \omega}) \right),
\end{eqnarray}
where the constants $B$, $C$, and $Z$ are defined in Eq. (\ref{BZC}),
$k$ is the wave vector along the stripe, $\omega$ is the frequency, the
``+'' sign corresponds to the symmetric, the ``--'' sign corresponds to
the antisymmetric deformations of the stripe, and
\begin{equation} \label{l01}
  \lambda_0 = - {\ep B (1 - e^{- \ls_s \sqrt{C}})\over 2 Z \sqrt{C}}.
\end{equation}
Similarly, for the spots we will have the following dispersion relation
\begin{eqnarray} \label{d2}
  i \al \omega && + {\ep^2 m^2 \over \rs_s^2} + \lambda_0 = \nonumber \\ 
  && - \ep B Z^{-1} I_m(\rs_s \sqrt{C + i \omega}) K_m(\rs_s \sqrt{C + i
    \omega}),
\end{eqnarray}
where $I_m$ and $K_m$ are the modified Bessel functions, $m$ is an
integer corresponding to the $m$-th surface mode, and $\lambda_0$ in
this case is
\begin{equation} \label{l02}
  \lambda_0 = - {\ep^2 \over \rs_s^2} - \ep B Z^{-1} I_m( \rs_s
  \sqrt{C}) K_m( \rs_s \sqrt{C}).
\end{equation}

The instabilities occur when $Im~ \omega < 0$.  These transcendent
equations can be solved for $\ls_s$ and $\rs_s$, respectively, when
$Im~\omega = 0$, for given $k$ or $m$. Then, using the dependences shown
in Fig. \ref{alar}, one can find the critical values of $A$ at which
different types of the instabilities occur.

Let us consider the case of the fast inhibitor $\al \gg \ep$. Then,
according to Eq. (\ref{d2}), at $A = A_{c2}^{(2)} = -0.56$ the spot will
become unstable with respect to the $m = 2$ mode, which corresponds to a
dumbbell-shaped deformation. Also, from Eq. (\ref{d2}) follows that the
spot destabilizes with respect to the $m = 0$ mode if $A < A_b^{(2)} =
-0.72$. This is in perfect agreement with the results of the
simulations.

Similarly, according to Eq. (\ref{d1}), the stripe becomes unstable with
respect to antisymmetric fluctuations (wriggling) at $A > A_{c2}^{(1)} =
-0.61$, and with respect to symmetric fluctuations (corrugation) at $A >
A_{c1}^{(1)} = -0.32$. The minimum width of the stripe is determined by
the overlap of the fluctuations of the activator, so it is not taken
into account in Eq. (\ref{d1}) \cite{ko:ufn89,ko:book,mo1:pre96}. If we
do take it into account, we will obtain that the stripe is unstable at
$A < A_b^{(1)} = -0.74$. So, here the agreement between the predictions
of the theory and the simulations is excellent as well.

The type of the complex pattern that forms in the late stages of the
destabilization of the simple patterns is determined by the dynamics of
its interface. For description of pattern dynamics in higher-dimensional
N systems Ohta, Mimura and Kobayashi developed an approach which allowed
them to reduce the equations similar to Eqs. (\ref{1}) and (\ref{2}) to
the problem of the interface dynamics in the case of slow inhibitor in
the limit $\ep \rightarrow 0$ and analyzed the early stages of the
transverse instability development \cite{ohta89}. Goldstein, Muraki and
Petrich derived an equation of the interface dynamics for a simple
system of FitzHugh-Nagumo type in the limit of fast inhibitor and weak
activator-inhibitor coupling and showed that the destabilization of
simple patterns lead to the formation of the connected labyrinthine
patterns \cite{goldstein96}.  Muratov derived the general equation of
the interface dynamics for N systems described by Eqs. (\ref{1}) and
(\ref{2}) and showed that in the limit $\ep \rightarrow 0$ and $\al \gg
\ep$ only multidomain patters must form as a result of the instability
and self-replication of simple patterns \cite{m:pre96}. However, because
of the slow dependences of certain parameters on $\ep$, multidomain
patterns should in fact form only at $\ep \lesssim 0.01$ in the
considered system \cite{m:pre96}. Yet the interfacial approach remains a
good approximation for the dynamics of the pattern for $\ep = 0.05$. In
this sense the region $0.01 \lesssim \ep \ll 1$ can be considered a
``crossover'' region between the labyrinthine and the multidomain
patterns. This is the reason why we see both connected and disconnected
labyrinthine patterns in our simulations. When $A$ is not far from
$A_{c2}^{(2)}$ the transverse instability is not very strong, so
connected labyrinthine patterns form (Fig. \ref{nosplit}). Here the
stripe shape is more favorable than the spot shape. As was noticed in
the previous Section, when $A$ is close to $A_{c2}^{(2)}$ (in the
simulations we used $A = -0.50$), a spot destabilizes into a single
stripe which does not branch. This may be qualitatively explained by the
following argument. In order for branching to occur, a spot has to be
unstable with respect to the $m = 3$ mode. According to Eq.  (\ref{d2}),
this should happen at $A > A_{c3}^{(2)} = -0.45$. For $A_{c2}^{(2)} < A
< A_{c3}^{(2)}$ the stripe is in turn unstable with respect to
wriggling, so as a result of the instability of a spot for those values
of $A$ the wriggled stripe (Fig.  \ref{wrig}) will eventually form in
the system. This is precisely what we see in our simulations.

When the value of $A$ gets larger, the transverse instability gets
stronger and the spot shape becomes more favorable. There we see domain
splitting predicted in Ref. \cite{m:pre96} and, therefore, the {\em
  disconnected} labyrinthine pattern, which is the counterpart of the
multidomain pattern in this case. We emphasize that this will happen
only when $\ep$ is relatively large; according to our simulations,
indeed, for $\ep \leq 0.01$ only multidomain patterns form in the
system.

Another important thing about the complex domain patterns is that the
multidomain patterns may coexist with the labyrinthine patterns. As was
shown by Muratov, for the same values of the parameters $\ep$, $\al$ and
$A$ one could excite both multidomain and labyrinthine patterns by
choosing appropriate initial conditions \cite{m:pre96}. For example, one
could take the pattern that formed in the end of the simulation of Fig.
\ref{turing} and use it as an initial conditions for the run with the
value of $A$ corresponding to the stable homogeneous state. Then in the
course of the system's evolution the domains will shrink and some of
them will disappear, but in the end the system will be filled with the
multidomain pattern similar to the one in Fig. \ref{turing} ($t = 955$),
rather than with the labyrinthine pattern.

Recently, Hagberg and Meron studied numerically the formation of
labyrinthine patterns in a bistable N system and explained this effect
on the basis of the non-trivial properties of solitary fronts that form
only in bistable systems \cite{hagberg:chaos94}. However, according to
the experimental observations \cite{lee:pre95} and our numerical
simulations, labyrinthine patterns may form both in monostable and
bistable systems. In order to make the system of Eqs. (\ref{act}) and
(\ref{inh}) bistable one needs to add a coefficient $\gamma$ in front of
$\th$ on the right-hand side of Eq. (\ref{inh}). Then for $\gamma = 1$
the system will be monostable, whereas for smaller $\gamma$, for example
$\gamma = 0.5$, the system is bistable. We did not see any qualitative
difference between the patterns in these two cases. In the monostable
systems the solitary fronts do not exist at all, so the domains always
have finite width at least in one direction. The properties of such
patterns are different from those of the solitary fronts, and are
essentially determined by the non-local interaction of different
portions of the pattern's interfaces
\cite{ko:ufn89,ko:book,mo1:pre96,m:pre96}.  Besides, in the bistable
systems with $\gamma \gg \ep$ and $\ep \ll 1$ the solitary fronts are
always unstable with respect to the transverse instability. Indeed,
according to Eq.  (\ref{d1}) with $B = 4 \gamma$, $C = 1 + \frac{1}{2}
\gamma$, and $\ls_s = \infty$, the front is unstable with respect to the
transverse perturbations with the wave vector $k \sim \ep^{-1/3}$. So,
in general one cannot use the properties of the solitary fronts to
explain the formation of complex domain patterns in N systems. Notice
that according to the similar argument, any pattern whose characteristic
size is much greater than $\ep^{1/3}$ is unstable with respect to
transverse perturbations both in the monostable and bistable systems.

Let us now consider the case of slow inhibitor $\al \lesssim \ep$. Since
the prevailing shape in the simulations in this case is a spot, we will
look for the instabilities of the circular domain. We solved Eq.
(\ref{d2}) in the case $\al \lesssim \ep$ for $m = 0, 1$ and 2. We found
qualitative agreement with the results of the general asymptotic theory
for instabilities of domain patterns \cite{mo1:pre96}.

For $m = 0$ the instability occurs at $Re~\omega \not= 0$. This
instability leads to the transformation of the static spot into a
radially-symmetric pulsating (breathing) spot. Such pulsating spots
were observed in the numerical simulations and the experiments
\cite{ko:ufn90,ko:ufn89,ko:book,haim96,olkg87}. The instability occurs when
the radius of a spot $\rs_\omega^{min} < \rs_s < \rs_\omega^{max}$,
where $\rs_\omega^{min}$ and $\rs_\omega^{max}$ are the functions of
$\al$.  For $\ep = 0.05$ and $\al \gtrsim 0.02$ the spot is stable in
the whole region of its existence. For $\al < 0.009$ the spot is
unstable for all $\rs_s$.

The $m = 1$ instability leads to the transformation of a static spot
into traveling. According to Eq. (\ref{d2}), a spot becomes unstable
with respect to the $m = 1$ mode when $\rs_s > \rs_T$, where $\rs_T$ is
a function of $\al$ and $A$. Notice that the general criterion of such
transformations was obtained by Osipov in Ref. \cite{o:pd96}. For the
same values of $A$ the $m = 1$ instability always happens at smaller
values of $\al$ than the $m = 0$ instability. The instabilities for $m
\geq 2$ with $Re~\omega \not= 0$ occur at even smaller values of $\al$,
when the spot is already unstable with respect to $m = 0$ and $m = 1$
modes.

The results of the analysis of the instabilities of simple shapes
(spots and stripes) and the results of the numerical simulations can
be presented on the diagram (Fig. \ref{bif}). This diagram shows the
domains of existence of different types of patterns in the $\al - A$
plane for $\ep = 0.05$ when the homogeneous state of the system is
stable. All the simulations points, which are marked by Roman letters
in Fig. \ref{bif}, were obtained by using localized initial
conditions. The vertical lines in Fig. \ref{bif} correspond to the
values of $A$ at which different instabilities of the domain shapes
occur, calculated from Eqs. (\ref{d1}) and (\ref{d2}). One can see
that these lines separate the regions in which the corresponding types
of static patterns are observed in the simulations. The letter ``s''
corresponds to the simulations in which the aperiodic relaxation was
observed. The upper dashed line separates the region in which any
initial condition relaxes aperiodically to one of the static patterns
from the region in which the relaxation becomes oscillatory. As was
expected \cite{ko:ufn89,ko:book}, the transition from the aperiodic to
the oscillatory relaxation occurs at $\al \sim \ep$. Above the upper
dashed line the form of the patterns is essentially independent of
$\al$; depending on the value of $A$ and the initial condition one can
see spots, stripes, wriggled stripes, multidomain patterns and
labyrinthine patterns (Figs. \ref{turing}, \ref{nosplit}, \ref{corr},
\ref{wrig}) . Of course, for $\ep = 0.05$ one should use special
initial conditions (not localized) to excite the multidomain patterns.

Below the upper dashed line but above the upper solid line the
relaxation of the initial excitation of the system results in the
formation of a stable static pattern, although a few pulsations and spot
splittings associated with them may occur at the beginning [Fig.
\ref{waves}(c)], so the resulting pattern is disconnected for all values
of $A > A_{c2}^{(2)}$. The simulations of this type are marked ``ps'' in
Fig. \ref{bif}. Notice that in this parameter region self-replication of
spots does not occur, but the fact that the inhibitor is slow makes the
domain splitting easier, since the inhibitor lags behind the motion of
the interface driven by the transverse instability of the spot
\cite{m:pre96}.

The upper solid line represents the solution of Eq. (\ref{d2}) for $m =
0$ and shows the instability line for a spot with respect to pulsations
(breathing). The simulations show that below this line different dynamic
patterns form. When $A$ is big enough and $\al$ is just slightly below
the upper solid line, spot replication leading to the formation of
static or synchronously pulsating pattern is observed (Fig.
\ref{split}). These simulations are marked by ``p'' in Fig. \ref{bif}.
As was already mentioned above, for $\ep = 0.05$ self-replication does
not occur when the inhibitor is fast. Nevertheless, self-replication
does occur in the case of the slow inhibitor ($\al \lesssim \ep$). In
this case the transverse instability, which is the primary cause of the
domain splitting and self-replication \cite{m:pre96}, is assisted by the
instability which leads to the radially-symmetric pulsations. This is
the reason why spot replication does not occur above the upper solid
line, which is the pulsation instability threshold for a spot.

As can be seen from Fig. \ref{bif}, the dominant type of dynamic
patterns for $\al \sim \ep$ is the turbulence (Fig. \ref{turb}, the
points marked ``t'' in Fig. \ref{bif}). When the two spots come at
distances less or of order 1 to each other, the inhibitor may not be
able to suppress the growth of one spot due to the shrinkage of the
other (the activator repumping effect) what may result in the
disappearance of one of the spots. However, the surviving spot may
self-replicate in turn and create another spot. Also, if a spot does not
have other spots around, it may transform into an annulus as a result of
the local breakdown in the spot's center, and the annulus may then break
up into a number of spots as a result of the transverse instability. It
is these three processes uncorrelated in space that make the turbulence
possible in the K$\Omega$N systems.

The turbulence is observed at relatively large values of $A$. This is
not surprising. Since the turbulence is caused by the self-replication
process, it may occur only when the spot is able to replicate, that is,
when $A > A_{c2}^{(2)}$.  For smaller values of $A$ the spots do not
self-replicate, instead they collapse after a few periods of pulsations
(simulations marked ``c'' in Fig. \ref{bif}). The meaning of the
separation between the region where the turbulence and the synchronously
pulsating patterns are realized is less obvious.  Qualitatively, the
disappearance of some of the domains in the course of the pattern's
dynamics and the local breakdown, the processes that cause the
stochastization, occur easier when the inhibitor is slower, that is,
when $\al$ is smaller. Of course, in order to make a quantitative
explanation of this separation, one has to solve a highly nonlinear
free-boundary problem in two dimensions.

All the dynamic patterns mentioned above are observed above the lower
solid line, which is the stability margin for a spot for $m = 1$
obtained from Eq. (\ref{d2}). Below this line a static spot
destabilizes and transforms to traveling. In this region only
autowaves (the simulations marked ``a'') form from a localized initial
perturbation [Fig. \ref{waves}(a)]. The autowave patterns that form
below the lower solid line are essentially the same for all values of
$\al$ (aside from the time and length scales of the pattern) and in
fact do not differ from the autowaves forming in $\Omega$N-systems
with $\ep \gtrsim 1$.  This is because at $\al \ll \ep$ the diffusive
precursor is not able to form in front of the traveling pattern front
\cite{ko:ufn89,ko:book}. Observe that according to Fig. \ref{bif}, no
complex static or dynamic patterns (except autowaves) form in the
system at any $A$ if $\al \lesssim \ep^2$.  This fact is in total
agreement with the general qualitative theory
\cite{ko:ufn90,ko:ufn89,ko:book} and with the conclusions of the
general asymptotic theory of instabilities \cite{mo1:pre96,o:pd96}.

Hagberg and Meron explained self-replication of spots and formation of
turbulence as the consequences of the parity-breaking bifurcations
[nonequilibrium Ising-Bloch (NIB) front transitions] of the planar
fronts in bistable N systems with the weak activator-inhibitor
coupling \cite{hagberg:chaos94}. Although their approach is useful for
the qualitative or heuristic explanation of the formation of the
dynamic patterns discussed above, it is highly inadequate for making
quantitative predictions in general. Indeed, the process of
self-replication observed in our simulation cannot be viewed as a
consequence of local NIB transitions (reversal of the propagation
direction of the portions of the spot's interface), since, according
to our numerical simulations, the {\em whole} spot's interface
reverses its propagation velocity in the course of self-replication
(Fig. \ref{sd}).  Furthermore, as can be seen from Fig. \ref{bif}, the
region of the system's parameters in which spot self-replication is
realized is determined by the instabilities of the static spot, and
not of the planar front, or the planar stripe, which is the
counterpart of the planar front for the monostable systems.  One can
show that according to Eq. (\ref{d1}), both the instability of the
stripe with respect to pulsations [the ``+'' sign in Eq. (\ref{d1})]
and the instability leading to the transformation of the static stripe
into traveling [the ``--'' sign in Eq. (\ref{d1})] lie considerably
lower than both solid lines in Fig. \ref{bif}, which correspond to the
respective instabilities of the spot. Hagberg and Meron predict domain
splitting and formation of disconnected labyrinthine patterns only
close to the NIB transitions (where $\al \sim \ep$)
\cite{hagberg:chaos94}, and yet domain splitting and the formation of
disconnected labyrinthine pattern occurs solely due to the transverse
instability far from the presumed NIB transitions
(Fig. \ref{corr}). Also, as was already mentioned, the turbulence that
was observed in our simulations is different from the spiral
turbulence observed by Hagberg and Meron in the bistable system with
relatively weak activator-inhibitor coupling
\cite{hagberg:prl94,hagberg:chaos94}.  One could think of the
turbulence observed by Hagberg and Meron as intermediate between the
spiral turbulence observed in the excitable autowave media ($\Omega$N
systems) and the turbulence observed by us in a K$\Omega$N system. In
our simulations the nucleation of spiral vortex pairs is not allowed
by the local breakdown. The turbulence is produced by the constant
self-replication of spots with the stochastization caused by the
disappearance (annihilation) of some of the domains because of their
strong interaction with the neighbors, and the spontaneous creation of
new interfaces (transformation of a spot into an annulus) due to the
local breakdown, which occurs, if the size of the domain becomes big
enough (Fig. \ref{turb}). All this suggests that in the general N
systems with $\ep \ll 1$ strong non-local interaction between the
different portions of the domain interfaces and between different
domains, high curvature of the domain interfaces, the time lag between
the motion of the interface and the reaction of the inhibitor, and the
process of local breakdown are crucial and in fact determine the type
of the pattern that will form in the system from a localized stimulus
for the given values of the system's parameters.

Before concluding this Section, let us discuss how the changes in the
values of $\ep$ should affect the bifurcation sequences and the pattern
formation scenarios in the considered systems. As was already mentioned,
the value $\ep = 0.05$ corresponds to a ``crossover'' between the
asymptotically small values of $\ep$, and the relatively large $\ep \sim
1$. The value of $\ep = 0.05$ is reasonably small to admit strong
separation of the length scales of the activator and the inhibitor, yet
it is not very small in the asymptotic sense, for which in the
considered model we should have $\ep \lesssim 0.01$. It is clear that if
the value of $\ep$ is increased, the transverse instabilities will
become weaker, so the vertical lines corresponding to the instabilities
of a spot in Fig. \ref{bif} will move to the greater values of $A$. For
these values of $\ep$ and $\al \gg \ep$ the stripe shape will become
dominant over the spot shape, so the typical complex pattern forming in
the system will be the labyrinthine pattern which consists of long
wriggled stripes, which may still be disconnected [Fig. \ref{waves}(d)].
This will also be true for the Turing patterns forming as a result of
the instability of the homogeneous state.  If the value of $\al$ gets
smaller, the turbulent patterns will form. Our simulations show that in
this situation the pattern's oscillations are less likely to synchronize
than in the case of smaller $\ep$ because of the stronger
stochastization due to the local breakdown, so it would be easier for
the turbulent patterns to form. If $\ep$ is even greater, the patterns
can no longer be viewed as having sharp interfaces, so the entire
phenomenology of pattern formation will change.

On the other hand, if the value of $\ep$ is decreased, the transverse
instability will become stronger, and the dominant shape will become the
spot shape. In this case all vertical lines which correspond to the
instabilities of a spot in Fig. \ref{bif} will move toward $A = A_b$. In
the case $\al \gg \ep$ only disconnected labyrinthine patterns or
multidomain patterns will form as a result of the transverse instability
\cite{m:pre96}. The characteristic size of such patterns will be
$\ep^{1/3}$ \cite{mo1:pre96,m:pre96}, so at very small $\ep$ any spot
will interact with many other spots. Because of this interaction, for
$\al \lesssim \ep$ the pulsating patterns will tend to synchronize more,
so the synchronously pulsating multidomain patterns should exist in a
wider range of the system's parameters. Also, as follows from the
general asymptotic theory of instabilities \cite{mo1:pre96}, for
extremely small $\ep$ ($\ep \lesssim 10^{-6}$) the instability of a spot
with respect to the fluctuations leading to the formation of a traveling
spot will always occur at larger values of $\al$ than the pulsating
instability, so it is natural to conclude that complex dynamic patterns,
such as pulsating multidomain patterns and turbulent patterns, cannot be
excited by a localized stimulus at such small values of $\ep$.

\section{conclusion}

In this paper we performed a complete numerical study of different
types of domain patterns in a two-dimensional N system and
investigated all major scenarios of their formation. We confirmed the
conclusions of Kerner and Osipov \cite{ko:ufn90,ko:ufn89,ko:book} that
the type of the domain patterns forming in the N systems is determined
mainly by the two basic parameters of the system: $\al = \tau_\th /
\tau_\et$ and $\ep = l / L = \sqrt{D_\th \tau_\th / D_\et
\tau_\et}$. These parameters are the ratios of the characteristic time
and length scales of the variation of the activator and the inhibitor
and are determined by the local kinetic coefficients: the relaxation
times and the diffusion coefficients $D_\th$ and $D_\et$. In real
systems these kinetic coefficients strongly depend on the excitation
level of the system and the state of the environment, the presence of
small amounts of impurities or catalysts which, for example, may
change the rates of recombination of nonequilibrium carriers in
semiconductors or the rates of chemical reactions, and so on. For
example, by varying only the temperature of the semiconductor lattice,
one can significantly change the critical parameters of the
electron-hole plasma described by Eqs. (\ref{burn:i}) and
(\ref{burn:a}) \cite{ko:jetp85}.

We have shown that by changing the values of $\ep$ and $\al$ and the
control parameter $A$ (in the physical systems, such as electron-hole
plasma, $A$ is the system's excitation level), that is, in essentially
the same system the whole variety of domain patterns and pattern
formation scenarios is realized. This general conclusion explains
theoretically the results of recent experiments by Lee {\em et al.} on
the FIS reaction \cite{lee:sci93,lee:ln94,lee:pre95} where they showed
that in this chemical system by changing relatively weakly its chemical
composition and the form of the initial perturbation it is possible to
excite all major types of domain patterns and see various scenarios of
their formation, which are qualitatively the same as those observed in
our simulations.

In KN systems, that is, when $\ep \ll 1$ and $\al \gg \ep$ only static
patterns form. When the homogeneous state of the KN system is stable,
these patterns can be excited by applying a sufficiently strong
perturbation (hard excitation). We found that by changing only the
parameter $A$ and the form of the initial perturbation one can excite
localized spots and stripes, connected (Fig. \ref{nosplit}) and
disconnected (Fig. \ref{corr}) labyrinthine patterns, wriggled stripes
(Fig. \ref{wrig}) and multidomain patterns (Fig. \ref{turing}). All
these patterns were found in the chemical experiments
\cite{lengyel93,kapral,lee:pre95}. Ordered multidomain patterns were
also observed in the high frequency gas-discharge experiments
\cite{ammelt93} and in the combustion experiments \cite{gorman94:order}.
We also found that at the same parameters of the KN system it is
possible to excite a great variety of shapes of the static patterns by
changing only the form of the initial perturbation. In particular, it is
possible to excite the labyrinthine and multidomain patterns in the
system with the same values of the parameters $\ep$, $\al$, and $A$
\cite{m:pre96}. This variety of domain shapes is observed when $\ep$ is
not very small, since for very small values of $\ep$ only disconnected
multidomain patterns will form \cite{m:pre96}. However, such small
values of $\ep$ can hardly be realized in a typical experimental
situation. That's why in our paper we paid particular attention to the
case $\ep = 0.05$.

Static domain patterns in KN systems may also form spontaneously as a
result of Turing instability of the homogeneous state. In the ideally
homogeneous KN systems these patterns are as a rule quasiperiodic (Fig.
\ref{turing}) and in general their period is not the same as the period
of the critical fluctuation $2 \pi / k_0$ [see Eq. (\ref{Ac})], but is
determined by the stability of the pattern.  In real systems the type of
the pattern will be determined by the small local inhomogeneities.
Depending on the form of the inhomogeneity, and also on the system's
parameters, all types of static domain patterns will form spontaneously
in KN systems.

In $\Omega$N systems, that is when $\al \ll 1$ and $\al \lesssim
\ep^2$, only the uniform relaxation self-oscillations may form
spontaneously. In such systems with the stable homogeneous state
various autowave patterns, including expanding traveling waves
[Fig. \ref{waves}(a)] and spiral waves, can be excited by an external
perturbation. If the initial conditions are sufficiently random, the
spiral turbulence [Fig.  \ref{waves}(b)] forms in $\Omega$N
systems. These patterns were first discovered by Zaikin and
Zhabotinsky in an oscillatory chemical reaction \cite{zaikin70} and
have been subsequently studied for more than two decades in a variety
of systems \cite{field,vasiliev,mikhailov,kapral}. Notice that the
autowave patterns are also observed in the FIS reaction
\cite{lee:pre95}.

The most diverse picture of pattern formation is observed in the case of
K$\Omega$N systems, that is, when $\ep \ll 1$ and $\ep^2 \lesssim \al
\lesssim \ep$. In these systems both the uniform self-oscillations and
the Turing patterns may form spontaneously. In the K$\Omega$N systems
with the stable homogeneous state it is possible to excite static,
traveling, and pulsating (breathing) patterns. We showed that in these
systems the remarkable effect of self-replication of spots (Figs.
\ref{split} and \ref{sd}) recently discovered in the same FIS reaction
\cite{lee:ln94,lee:pre95} is realized. Depending on the system's
parameters, this process may lead to the formation of static or
pulsating multidomain pattern, or to the formation of turbulence which
is qualitatively different from the spiral turbulence observed in
$\Omega$N systems. This turbulence consists of random creation and
annihilation of spots. Precisely this kind of turbulence was observed in
the experiments in the very same FIS reaction \cite{lee:pre95}, and also
in the combustion experiments \cite{gorman94:chaos}. 

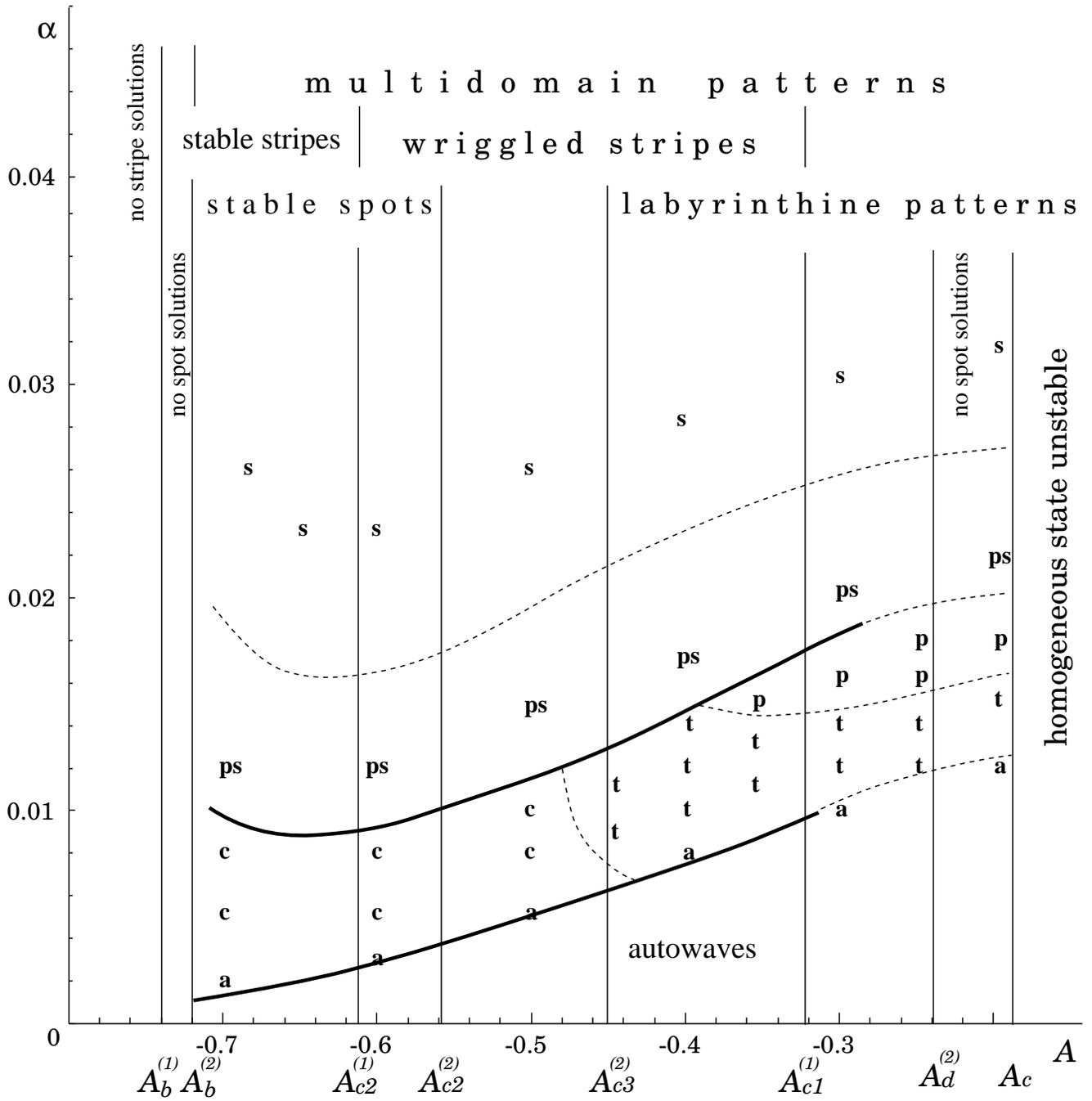
\begin{figure}
\caption{ \label{bif}
  The domains of existence of different patterns at $\ep = 0.05$. The
  upper part of the figure shows the regions where the corresponding
  patterns may exist at $\al \gg \ep$. The lower-case letters indicate
  the long-time behavior of the system at the parameters corresponding
  to the position of the letter when the system is locally excited at $t
  = 0$: ``s'' --- the system aperiodically relaxes to a static pattern;
  ``ps'' --- the system relaxes to a static pattern, but the relaxation
  has oscillating character, a few splittings may occur at the
  beginning; ``p'' --- as a result of self-replications at the beginning
  a stationary pulsating (breathing) pattern forms in the system; ``c''
  --- the initial excitation collapses as a result of the growing
  amplitude of pulsations; ``t'' --- turbulence develops in the system;
  ``a'' --- the initial excitation transforms into an autowave traveling
  through the system and disappearing at the boundaries. The upper
  dashed line shows schematically the region where the character of
  pattern's relaxation changes from aperiodic to oscillating. The upper
  solid line is the threshold of the instability with respect to the
  uniform pulsations for a radially symmetric AS, obtained from Eq.
  (\protect\ref{d2}). The lower solid line is the threshold of the
  instability leading to the transformation of the radially symmetric AS
  into traveling, obtained from Eq. (\protect\ref{d2}). The lower dashed
  lines show schematically the borders of the parameter regions where
  pulsating patterns or turbulence are realized.  }
\end{figure}

We would like to emphasize that the whole variety of pattern formation
scenarios observed in our numerical simulations and illustrated in Fig.
\ref{bif} is explained with the remarkable accuracy by the asymptotic
theory of instabilities of domain patterns in reaction-diffusion systems
\cite{mo1:pre96}. This is because the scenarios of pattern formation are
determined by the two processes: at first any initial perturbation
quickly transforms to a state close to some stationary state, then the
evolution of that quasistationary state is determined by its
instabilities and the form of the critical fluctuations with respect to
which the instability is realized. Indeed, the stationary states
correspond to the solutions of Eqs. (\ref{1}) and (\ref{2}) with the
right-hand sides equal to zero, so when we use the initial conditions
which are significantly different from any stationary state, we in fact
make the right-hand sides of Eqs. (\ref{1}) and (\ref{2}) large and,
therefore, the time derivatives of $\th$ and $\et$. The large time
derivatives will cause such changes in $\th$ and $\et$ which will lead
to the transformation of the initial condition to the state close to
some stationary state, in which the time derivatives of $\th$ and $\et$
are small. For example, if a square of relatively small size is used as
the initial condition, it will first transform to a state close to a
spot; if a square of large size ($\gg L$) is used, it will first
transform into an annulus as a result of the local breakdown in the
center of the square (this effect was studied in detail in Refs.
\cite{ko:ufn90,ko:ufn89,ko:book}). If the system is bistable, the
initial condition in the form of a large square may trigger the wave of
switching from one stable homogeneous state to the other. The evolution
of these states close the stationary states, and, therefore, the pattern
formation scenarios, will be determined by their stability: a spot may
form from a square of small size if the parameters of the system are
such that it is stable, or an annulus may form from a square of large
size. The radius of the spot, or the width and size of the annulus will
be determined by the corresponding stable states and will only depend on
the parameters of the system (such as $\ep$ and $A$). If the parameters
of the system are such that these states happen to be unstable, then,
depending on the type of the instability, which is determined by the
parameters $\ep$, $\al$, and $A$, all kinds of complex patterns will
form.

One of the bright pattern formation scenarios consists of the
transformation of the localized excitations into the patterns that fill
the entire system. In certain sense one could think of this as the
self-completion of a pattern from its small fragment. According to Figs.
\ref{turing} -- \ref{wrig}, the form of such patterns depends on both
the initial conditions and the integral parameters of the system (in our
case $\ep$, $\al$, and $A$) which are determined by the system's
kinetics. It is important that by changing only these integral
parameters it is possible to excite qualitatively different patterns for
the same initial condition. Thus, the systems we consider have a
remarkable property --- a kind of associative memory (see also Refs.
\cite{ko:ufn89,ko:book}): the form of a pattern is determined by the
integral parameters of the {\em ideally homogeneous system}, and they
can be reconstructed with certain probability from a small fragment
(sufficiently localized initial perturbation).

Patterns of the same morphology as those in the KN system studied by us
also form in a variety of the equilibrium systems, such as garnet
ferromagnets, ferroelectric and ferrofluid films, Langmuir monolayers,
phase-separating copolymer blends (for a recent review and the
references see Ref. \cite{seul95}). Our simulations suggest that the
complex domain patterns, such as multidomain or labyrinthine patterns,
are driven by the dynamics of their interfaces coupled to the
long-ranged inhibitor field. It is, therefore, natural to expect
qualitatively the same pattern formation scenarios in the equilibrium
systems with the competing repulsive interactions and the strong
separation of length scales. In the nonequilibrium systems the inhibitor
does not necessarily react on the motion of the pattern interfaces
instantaneously. The time lag of the inhibitor makes the existence of
complex dynamic patterns possible in the nonequilibrium systems.

The only kinds of patterns that form in the experiments with the cellular
flames \cite{gorman94:hopping} and with the gas-discharge system
\cite{ammelt93} we did not see in our numerical simulations are
traveling spots and hopping patterns. Recent work of Krischer and
Mikhailov suggested that a sufficiently strong global coupling, which is
absent in our model, might be needed to see these kinds of patterns
\cite{krischer94}. 

\bibliographystyle{prsty} 
\bibliography{../main}

\narrowtext

\begin{figure*}[htb]
\caption{The nullclines of Eqs. (\protect\ref{q}) and (\protect\ref{Q}).}
\label{null}
\end{figure*}

\begin{figure}
\caption{ \label{as} The simplest domain patterns: (a) traveling, (b)
  static, (c) pulsating one-dimensional autosolitons. }
\end{figure}

\begin{figure}
\caption{ \label{turing} Formation of static Turing pattern. The
  distributions of the activator at times. The parameters used: $\ep =
  0.05, \al = 0.5, A = -0.1$. The system's size is $20 \times 20$. }
\end{figure}

\begin{figure}
\caption{ \label{unstab} Formation of static Turing pattern in an
  oscillatory system. The distributions of the activator at different
  times. The parameters used: $\ep = 0.05, \al = 0.05, A = -0.1$. The
  system's size is $20 \times 20$. }
\end{figure}

\begin{figure}
\caption{ \label{nosplit} 
  Formation of a connected labyrinthine pattern. The distributions of
  the activator at different times. The parameters used: $\ep = 0.05,
  \al = 0.2, A = -0.4$. The system's size is $20 \times 20$. }
\end{figure}

\begin{figure}
\caption{ \label{corr} Formation of a disconnected labyrinthine
  pattern. The distributions of the activator at different times. The
  parameters used: $\ep = 0.05, \al = 0.2, A = -0.25$. The system's size
  is $20 \times 20$. }
\end{figure}

\begin{figure}
\caption{ \label{wrig}
  Formation of a wriggled stripe. The distributions of the activator at
  different times. The parameters used: $\ep = 0.05, \al = 0.2, A =
  -0.45$.  The system's size is $20 \times 20$.}
\end{figure}

\begin{figure}
\caption{ \label{split}
  Self-replicating spots. The distributions of the activator at
  different times. The parameters used: $\ep = 0.05, \al = 0.015, A =
  -0.4$. The system's size is $20 \times 20$. }
\end{figure}

\begin{figure} 
\caption{ \label{osc}
  The value of $\et$ in the center of the system as a function of time
  in the simulation with $\ep = 0.05, \al = 0.015, A = -0.3$ which
  resulted in the pulsating multidomain pattern. The system's size is
  $10 \times 10$. }
\end{figure}

\begin{figure} 
\caption{ \label{sd}
  A close-up of a self-replicating spot. Same simulation as in Fig.
  (\protect\ref{split}). The region shown is $8 \times 8$.}
\end{figure}

\begin{figure}
\caption{ \label{turb}
  The onset of turbulence. The distributions of the activator at
  different times. The parameters used: $\ep = 0.05, \al = 0.01, A =
  -0.4$. The system's size is $10 \times 10$. }
\end{figure}

\begin{figure} 
\caption{ \label{chaos}
  The value of $\et$ in the center of the system as a function of time
  for the simulation of Fig. \protect\ref{turb}. }
\end{figure}

\begin{figure}
\caption{ \label{waves}
  The distributions of the activator at different times for different
  processes: (a) formation of an autowave ($\ep = 0.05, \al = 0.007, A =
  -0.3$, the system is $20 \times 20$); (b) formation of spiral
  turbulence ($\ep = 0.25, \al = 0.02, A = -0.3$, the system is $100
  \times 100$); (c) stabilization of a pattern formed in the process of
  splitting of spots ($\ep =0.05, \al = 0.02, A = -0.3$ the system is
  $10 \times 10$); (d) formation of a complex pattern outside the
  asymptotic regime ($\ep = 0.2, \al = 1.0, A = -0.15$, the system's
  size is $40 \times 40$).}
\end{figure}

\end{multicols}

\end{document}